# Radio Resource Management for Dynamic Channel Borrowing Scheme in Wireless Networks

Mohammad Arif Hossain, Shakil Ahmed, and Mostafa Zaman Chowdhury
Department of Electrical and Electronic Engineering
Khulna University of Engineering & Technology, Khulna-9203, Bangladesh
E-mail: dihan.kuet@gmail.com, shakileee076@gmail.com, mzceee@yahoo.com

*Abstract*— Provisioning of Quality of Service (QoS) is the key concern for Radio Resource Management now-a-days. In this paper, an efficient dynamic channel borrowing architecture has been proposed that ensures better QoS. The proposed scheme lessens the problem of excessive overall call blocking probability without sacrificing bandwidth utilization. If a channel is borrowed from adjacent cells and causing interference, we also propose architecture that diminishes the interference problem. The numerical results show comparison between the proposed scheme and the conventional scheme before channel borrowing process. The results show a satisfactory performance that are in favor of the proposed scheme, in case of overall call blocking probability, bandwidth utilization and interference management.

## I. INTRODUCTION

To cope with the demand of modern wireless communication system, it is mandatory to make the maximum use of radio resources within limited bandwidth at each cell so that a user can have maximum Quality of Service (QoS) at all the time. In a network, there may be unoccupied channels in the cells. These unoccupied channels can be the hot cake to fulfill the need of the excessive users of the other cells, if a scheme can be developed so that maximum utilization of these channels can be ensured [1]-[3]. Being deep concern of the burning issue, we have proposed a scheme, where, if there is more traffic in a cell compared to number of channels, the cell has the opportunity to take required channels from other cells, if any, of course after interference management. In our scheme, the required number of channels of reference cell (i.e. the cell where traffic intensity is higher) will be borrowed from adjacent cells, having maximum number of unused available channels. We propose that for interference management, the reference cell will be bifurcated, named inner and outer part [4]. The required or available number of channels will be provided to the inner part users of the reference cell after borrowing for interference management.

In the previous time, fixed channel assignment (FCA) [5] and hybrid channel assignment (HCA) [6] have been proposed to utilize the bandwidth without considering interference management. A channel borrowing without locking (CBWL) scheme is proposed in [7]. In the scheme, the cell will borrow a channel from its adjacent cells when a call arrives but can not be served by a normal channel. If the borrowed channel is obtained, the cell will use it with reduced power so that the borrowed channel is not necessary to be locked.

In the proposed scheme, number of users greater than the number of channels of the cell can be accommodated by dynamically borrowing the channels from other adjacent cells. That means either reference cell will take the required channel ($N_{req}$) (if the available channels are greater than the required channels) or the available channel ($N_{av}$) (if the available channels are less than the required channels) from the adjacent cells.

The rest of this paper is organized as follows: Section II shows the proposed dynamic channel borrowing scheme with block diagram. Call blocking probability using the queuing analysis for the proposed scheme is shown in Section III. In Section IV, the numerical results for our proposed scheme are shown. Finally, conclusions are drawn in the last section.

## II. PROPOSED DYNAMIC CHANNEL BORROWING SCHEME

Contemporary and future wireless network are required to serve maximum users. As there is excessive traffic now-a-days, a dynamic channel borrowing scheme can be an effective way, where maximum bandwidth utilization with less overall call blocking probability and QoS are ensured.

### A. System model

We consider a cluster of seven cells where three types of reused frequencies named *A, B, C* are available. Fig. 1 shows the frequency allocation of cells before channel borrowing of the system. Assume that the cell 1 is the reference cell where the traffic intensity is higher than the six cells of the cluster. It is our deep contemplation that results, if a cell (say cell 2) is found with maximum number of unoccupied channels ($C_{muc}$), the reference cell will continuously borrow from that cell till the channels are available. After that if the reference cell requires more channels, and then it will borrow channels from the cell which has maximum number of unoccupied channels (say cell 3) at that instant, which reduces the complexity of the system with commendable performance. This is shown in Fig. 2. In the mean time, whenever interference management is required, the cell will be bifurcated and the borrowed channels will be provided to the inner part users of the reference cell. This is shown in Fig. 3. And it will go the next cell (say cell 3) that has different frequency than of the previous two cells (i.e. cell 1 and 2), if more required. This is also shown in Fig. 2.

## B. Borrowing Architecture

If all the channels of reference cell 1 are busy, immediately traffic arrives that needs $N_{req}$ channels in cell 1. Fig. 4 shows the search of $N_{req}$ in the adjacent cell 2, 4 and 6 or 3, 5 and 7. $N_{req}$ will be borrowed from the cell with the maximum unoccupied channels of same frequency.

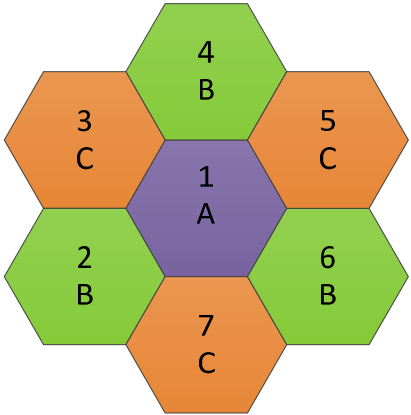

Fig.1. Frequency allocation before channel borrowing process.

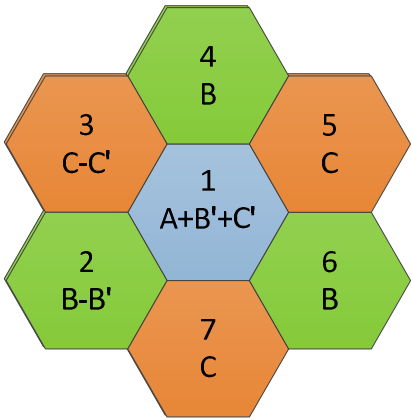

Fig. 2. Frequency allocation after channel borrowing process without interference management.

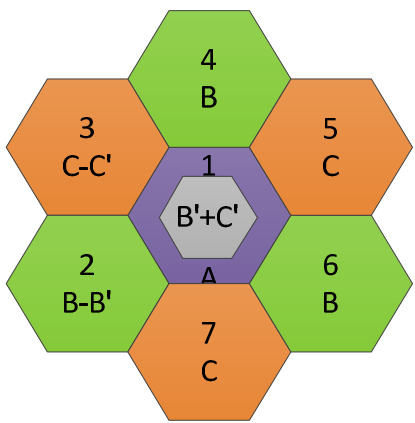

Fig. 3. Frequency allocation after channel borrowing process with interference management.

Fig. 5 shows the selection of the adjacent cell with maximum unoccupied channels. If cell 2 becomes the desired cell and $N_{req}$ is less than $N_{av,2}$ (the number of channels available in cell 2), then cell 1 will borrow $N_{req}$ channels and the total number of channel of cell 1 will be $N+N_{req}$. If $N_{req}$ is greater than $N_{av,2}$, then cell 1 will borrow the total unoccupied channels and the total number of channel of cell 1 will be $N+N_{av,2}$. If the total $N_{req}$ is not fulfilled, then cell 1 can borrow channel from the cell with the maximum number of unoccupied channels among cell 3, 5, and 7 (assume cell 3). If $N_{req(new)}$ ($N_{req(new)}= N_{req}- N_{av,2}$) is less than $N_{av,3}$ (the number of channels available in cell 3), then the total number of channel of cell 1 will be $N+N_{av,2}+N_{req(new)}$. If $N_{req(new)}$ is greater than $N_{av,3}$, then cell 1 will borrow $N_{av,3}$ channels and the total number of channels of cell 1 will be $N+N_{av,2}+N_{av,3}$.

## C. Interference Management

Fig. 2 shows that from cell 2 and 3, *B'* and *C'* are taken to reference cell 1, respectively. So, the bandwidth of cell 1 is increased from *A* to *(A+B'+C')*. But for *B'*, interference will occur for cell 4 and 6 and for *C'*, it is for cell 3 and 5 due to reused frequencies.

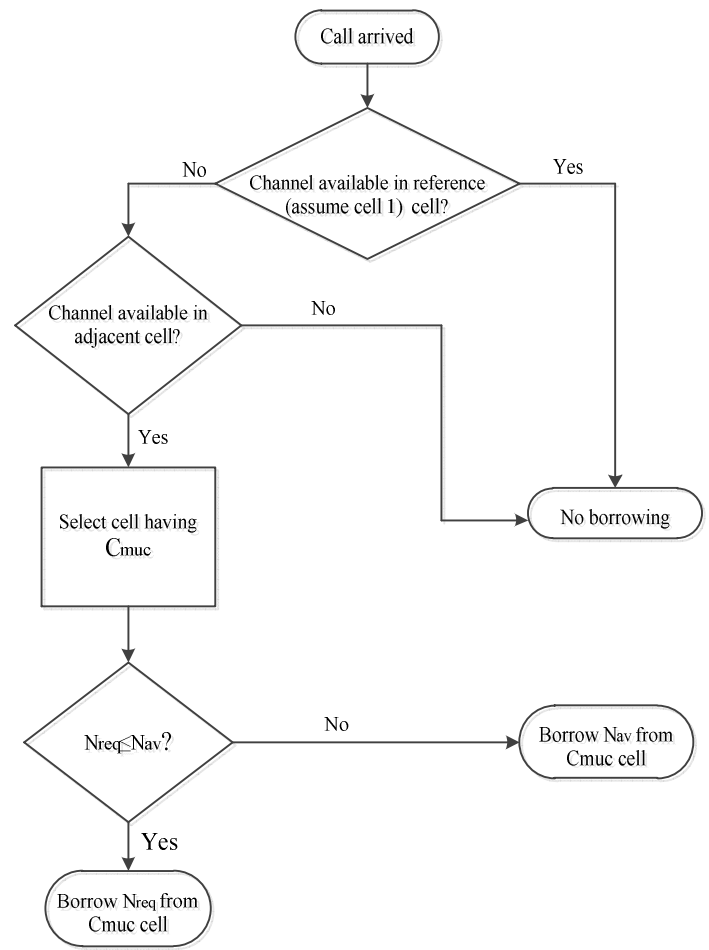

Fig. 4. Channel borrowing process

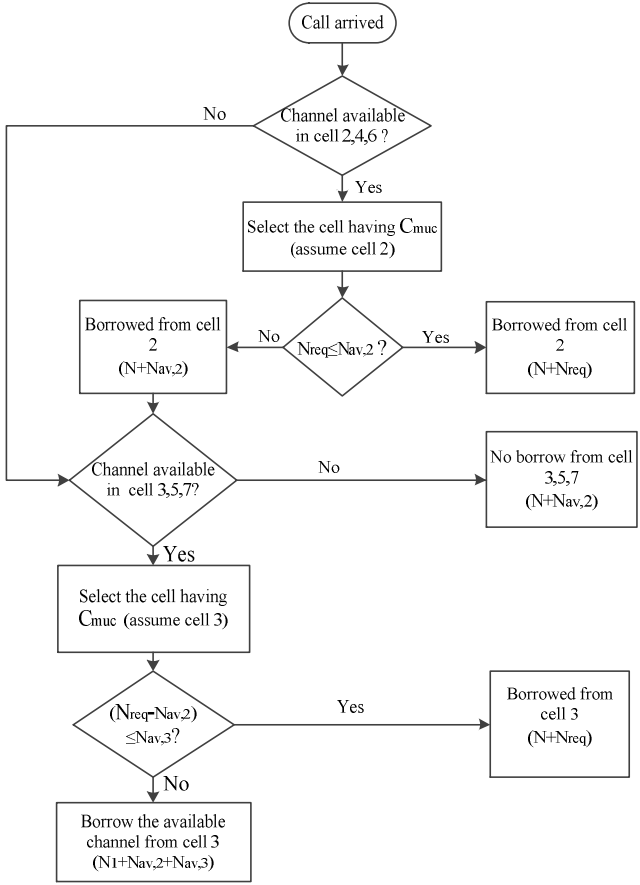

Fig. 5. Selection of adjacent cells for channel borrowing process.

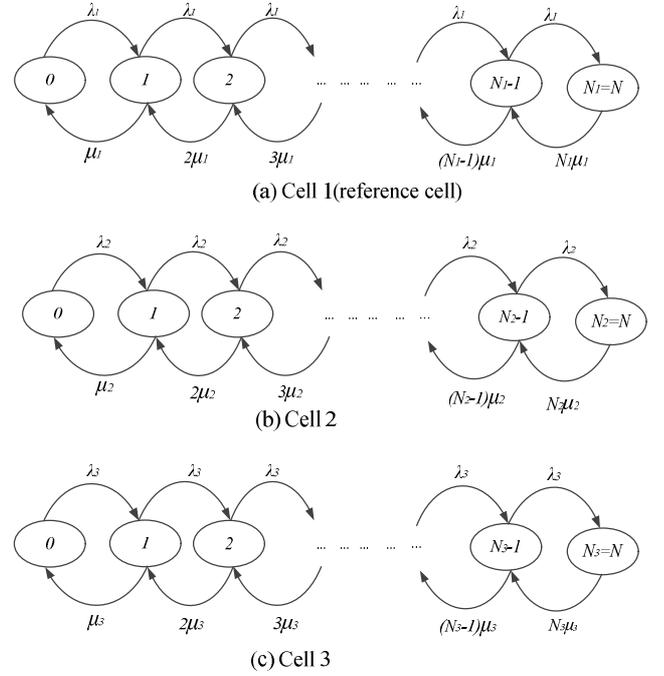

(a) Cell 1(reference cell)

(b) Cell 2

(c) Cell 3

Fig. 6. Markov chain for the proposed scheme before borrowing channels from cell 2 and cell 3.

Now for the management of interference, as shown in Fig. 3, the inner part users of cell 1 are provided with ($B'+C'$) and the outer part users are provided with rest of the resources.

In our proposed scheme, we use Okumura-Hata model for macro-cellular path calculation [8]. If the height of the base station be $h_b$, height of the mobile antenna be $h_m$, radius of the cell be d and $f_c$ be the carrier frequency then the Hata model shows a relation between them which is used for interference calculation.

$$L = 69.55 + 26.16 \log f_c - 13.82 \log h_b - a(h_m) + (44.9 - 6.55 \log h_b) \log d \quad (1)$$

where $a(h_m) = 1.1(\log f_c - 0.7)h_m - (1.56 \log f_c - 0.8)$

### III. QUEUING ANALYSIS

The proposed scheme can be modeled as $M/M/K/K$ queuing system and the call arriving processes are assumed to be Poisson.

The Markov Chain for the proposed scheme is shown in Fig. 6 and Fig. 7. From Fig. 6 and Fig. 7 it is clear that the arrival rate and departure rate is constant in each cell. Here, $N_1$ and $N_1'$ represent the total number of channels in cell 1 (Reference cell) before and after borrowing channels, respectively. Similarly, $N_2$ and $N_3$ represent the total number of channels in cell 2 and cell 3 before channel borrowing process, respectively. $N_2'$ and $N_3'$ represent the total number of channels in cell 2 and cell 3 after channel borrowing process, respectively. Besides, $\lambda_1$, $\lambda_2$ and $\lambda_3$ indicates the call arrival rate for cell 1,2 and 3, respectively and $\mu_1, \mu_2$ and $\mu_3$ indicates the channel holding time for cell 1,2 and 3, respectively.

Suppose, $M$ be the number of cells in a system and maximum number of calls that can be accommodated in a cell is $N_m$. Let, $P_m(i)$ be the steady state probability of the system in state i for cell m.

$$\sum_{i=0}^{N_m} P_m(i) = 1 \quad (2)$$

In the system, $\lambda_m$ and $\mu_m$ represent the call arrival rate and channel release rate for cell m, respectively.

$$P_m(i) = \frac{(\lambda_m)^i}{i!(\mu_m)^i} P_m(0), 0 \leq i \leq N_m \quad (3)$$

$$P_m(0) = \left[ \sum_{i=0}^{N_m} \frac{(\lambda_m)^i}{i!(\mu_m)^i} \right]^{-1} \quad (4)$$

Thus, from (2) to (4), the call blocking probability $P_{Bm}$ and overall call blocking probability of the system $P_{BT}$ can be calculated as per equations (5) and (6) shown below.

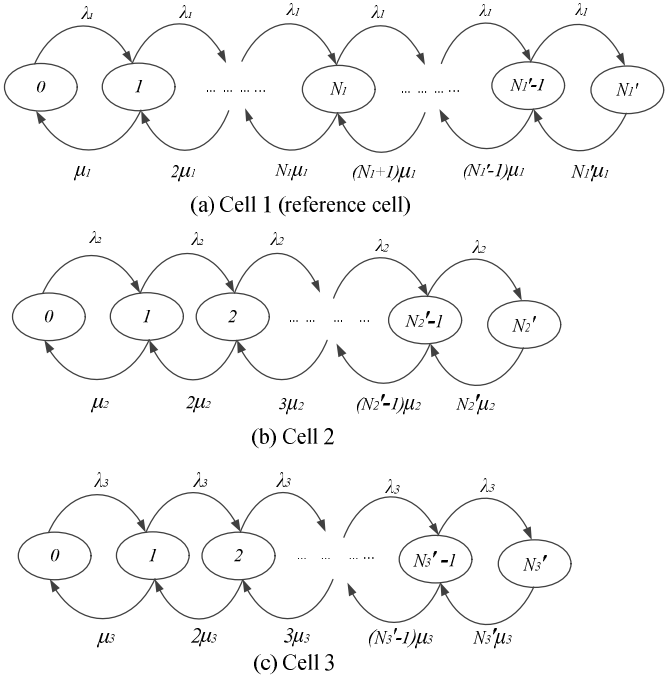

Fig. 7. Markov chain for the proposed scheme after borrowing channels from cell 2 and cell 3.

$$P_{Bm} = \frac{(\lambda_m)^{N_m}}{N_m!(\mu_m)^{N_m}} P_m(0) \quad (5)$$

$$P_{BT} = 1 - \frac{\sum_{m=1}^{M} \lambda_m (1 - P_{Bm})}{M \times \sum_{m=1}^{M} N_m} \quad (6)$$

## IV. NUMERICAL RESULT

In this section, we evaluate the performance of overall call blocking probability, overall bandwidth utilization and Signal to Noise Ratio (SINR) level of the proposed dynamic channel borrowing scheme with conventional scheme (i.e. without channel borrowing). We consider only the 1$^{st}$ and 2$^{nd}$ tier of the reference cell for interference management. Table 1 summarizes the values of the parameters that we used in our analysis.

Fig. 8 compares the overall call blocking probability between before and after borrowing channel. The figure describes the nobility of proposed scheme by showing less overall call blocking with compared to conventional scheme without borrowing channel.

Fig. 9 shows that the bandwidth utilization for the system based on the proposed scheme is maximized. The bandwidth utilization for channel without borrowing is poor. Moreover, the scheme without borrowing channel cannot maximize the bandwidth utilization especially after certain arrival rate.

TABLE 1: Summary of the parameter values used in analysis.

| Parameter | Value |
|---|---|
| No. of cell in each cluster | 7 |
| No of channel in each cell | 100 |
| No of reused frequency | 3 |
| Threshold value of channel for borrowing in each cell | 70 |
| Carrier frequency | 1800 MHz |
| Transmit signal power by the BS | 1.50 KW |
| Height of the BS | 100 m |
| Average channel holding time | 90 sec |
| Cell radius | 1 Km |

Fig. 10 shows the SINR levels of the proposed scheme and the conventional scheme without borrowing channels considering interference signals from adjacent cells and neighboring cells. The result implies that the SINR level decreases with the increment of the distance between the base station (BS) and the user of the reference cell.

The SINR level becomes insignificant when the user is far away from BS. By interference management, the SINR level can be increased which is very significant in case of channel borrowing architecture in comparison with the SINR level without interference management.

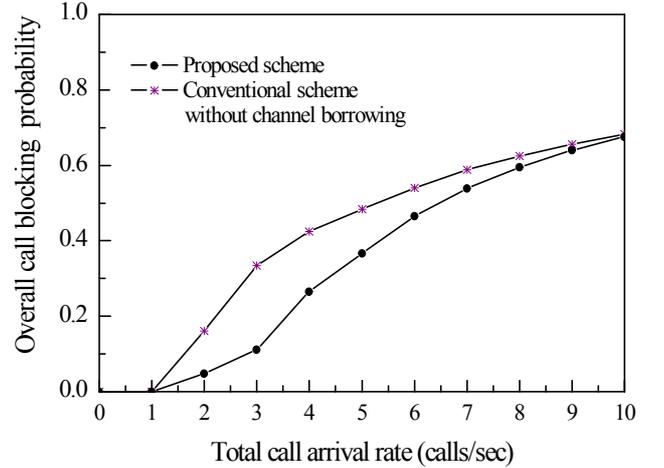

Fig. 8. Comparison of overall call blocking probability

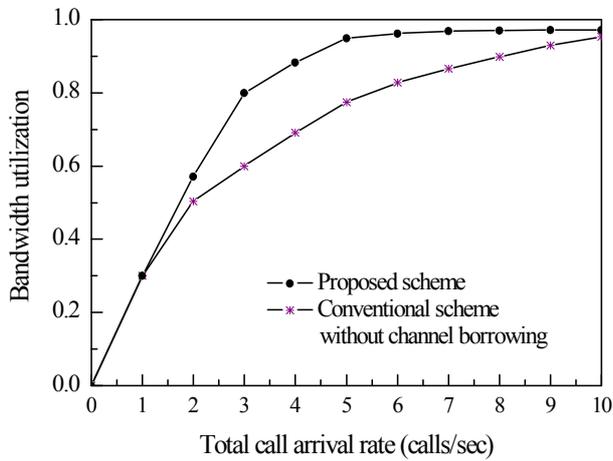

Fig. 9. Comparison of bandwidth utilization.

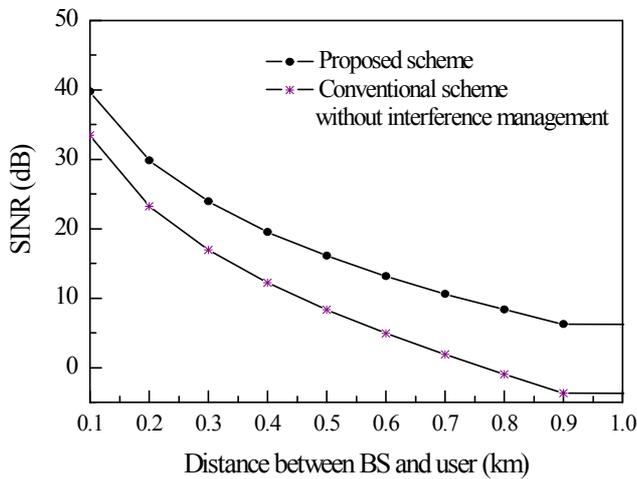

Fig. 10. Comparison of SNIR levels in case of interference management.

## V. CONCLUSION

Our deep scrutiny and simulation results show that the proposed scheme is quite operational with reduced overall call blocking probability without sacrificing bandwidth utilization as compared to conventional scheme (i.e. without channel borrowing). The proposed scheme also does the interference management which makes it more attractive. The proposed scheme is expected to be of considerable interest for future wireless communication. It is in our research to add priority scheme, multiclass traffic, MIMO in our future work.

REFERENCES

[1] Mostafa Zaman Chowdhury, Yeong Min Jang and Zygmunt J. Haas, "Call Admission Control Based on Adaptive Bandwidth Allocation for Wireless Networks," *IEEE/KICS Journal of Communications and Networks*, vol. 15, no. 1, February 2013.

[2] Ibrahim Habib, Mahmoud Sherif, Mahmoud Naghshineh and Parviz Kermani, "An Adaptive Quality Of Service Channel Borrowing Algorithm For Cellular Networks," *International Journal of Communication System*, vol. 16, no. 8, pp. 759–777, October 2003.

[3] Do Huu Tri, Vu Duy Loi and Ha Manh Dao "Improved Frequency Channel Borrowing And Locking Algorithm In Cellular Mobile Systems," In Proceeding of *IEEE International Conference on Advanced Communication Technology,* February 2009, pp. 214 – 217.

[4] Mostafa Zaman Chowdhury, Yeong Min Jang and Zygmunt J. Haas, "Cost-Effective Frequency Planning for Capacity Enhancement of Femtocellular Networks," *Wireless Personal Communications,* vol. 60, pp. 83–104, September 2011.

[5] Zuoying Xu, Pitu B. Mirchandani and Susan H. Xu "Virtually Fixed Channel Assignment In Cellular Mobile Networks With Recall And Handoffs," *Telecommunication Systems,* vol. 13, no. 2-4, pp. 413-439, July 2000.

[6] Shruti Pancholi M. Tech and Pankaj Shukla "Hybrid Channel Allocation in Wireless Cellular Networks," *International Journal of Communication Network Security,* vol. 1, no. 4, pp. 51-54, 2012.

[7] Hua Jiang and Stephen S. Rappaport, "CBWL: A New Channel Assignment and Sharing Scheme for Cellular Communication Systems," *IEEE Transactions on Vehicular Technology,* vol. 43, pp. 313-322, May 1994.

[8] Mostafa Zaman Chowdhury, Yeong Min Jang, Choong Sub Ji, Sunwoong Choi, Hongseok Jeon, Junghoon Jee, and Changmin Park, "Interface Selection For Power Management In UMTS/WLAN Overlaying Network," In Proceeding of *IEEE International Conference on Advanced Communication Technology*, February 2009, pp. 795-799.